# XIV ENCONTRO NACIONAL DE FÍSICA DA MATÉRIA CONDENSADA

HOLOPROJECTION OF IMAGES BY A DOUBLE DIFFRACTION PROCESS
J. J. Lunazzi
UNICAMP - Institut of Physics
C.P. 6165 - 13081 - Campinas-SP

Introduction

An image is said to be holographic when, after registering, it can be seen in three dimensions without using any special glasses or filters and continues parallax may be seen when looking around it. Holographic images are always registered by means of laser light, but may be seen under white light by applying conversion techniques.
In the case of the technique developed by S. Benton, only horizontal parallax is preserved, just the necessary for the horizontal position of the observer's eyes.
We demonstrate in this paper that the concept of a "holoimage" may be independent of the "graphic" ability corresponding to a registering material. This is due to new possibilities on the projection of images by means of difractive elements.
The three-dimensional distribution of white light may now be projected through a very small optical element, allowing for the direct creation of images that look like a Benton hologram but are phantasmagoric projections of the object itself.
Since the three-dimensional distribution of light from an object may be reproduced by a hologram, the system we named "holoprojector" also allows for the enlarging of a white-light hologram, reducing its weight and cost by orders of magnitude.
A third and very natural possibility is that of a common laser-light hologram, made by the off-axis technique of Leith and Upatnieks, being simultaneously enlarged and converted to white-light observation by a simple projection on a holographic screen.

Description

The chromatic encoding of views from an object was demonstrate to be a natural property of holograms (1) and diffraction gratings (2) by ray tracing calculations.
As a natural sequence of reasoning, we can show the decoding of this kind of images by projecting them onto a diffraction grating.
Figure 1 represents an object illuminated under white light represented by the light-spreading points 1-5. A diffraction grating R.D.. is interposed between the object and a lens L.

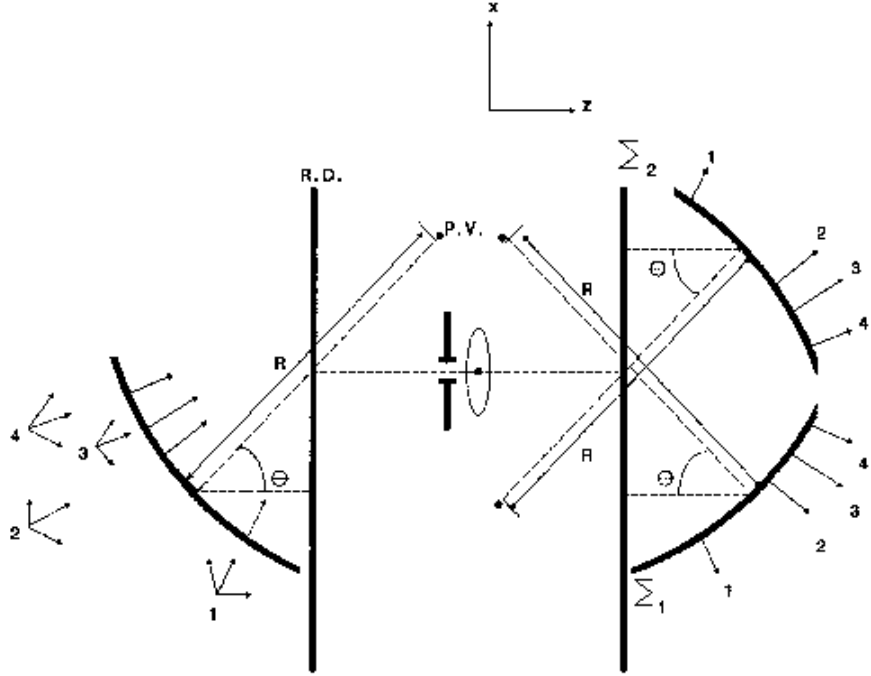

Figure 1

The aperture of this lens is very small, and we shall consider it as acting like a pinhole camera, for simplicity. The center 0 of the lens represents the viewpoint of the camera. From it we determined the viewpoint P. V. which is at the same distance from the point where the optical axis meets the grating, but at an angle θ that corresponds to the diffraction angle of a typical wavelenght value λ for light impinging perpendiculary to the grating. The location of this point is not arbitrary, since it allows for the determination of all the rays of wavelenght λ that, leaving the object points, will reach the point 0.

We may collect this rays at distance R from P. V. as the wave-front components of a spherical wave converging towards point P. V. The amplitude of this wave being described by A(x), its phase its constituted of a randomic term φ(t) which is different at each point on the wave due to the total incoherence of the light.

By using the paraxial approximation, the wave impinging on the grating may be expressed as:

$$\Psi_0(x) = A(x) e^{i\frac{kx^2}{2z}} e^{ikx \cdot \mathrm{sen}\,\theta} e^{i\phi(t)} \qquad (1)$$

with $k = 2\pi/\lambda$

We assume that the diffraction grating has a transmission amplitude of simple periodicity:

$$t(x) = a + b\cos\left(\frac{2\pi x}{d}\right) \qquad (2)$$

a, b being constants and d being the grating period.

After traversing the grating, the wave amplitude expression (1) must be multiplied by expression (2), giving two terms from which we separate the one that corresponds to propagation toward the lens L:

$$t(x)\Psi_0(x) \propto A(x)e^{if(t)}e^{i\frac{kx^2}{2z}}e^{i2px\left(\frac{\operatorname{sen}q}{l}-\frac{1}{d}\right)} \qquad (3)$$

We can then see that, for the particular value $\lambda = d \operatorname{sen}\theta$ the result is a spherical wave that converges toward the point 0. We neglected the fact that the amplitude variation A(x) may give its Fourier transform spectrum at this location.

By extending this procedure to any wavelength within the white light spectrum, we obtain a continous sequence of viewpoints like P. V. being unified at the single point 0.

This wavelength-encoded information may be easily decoded by just replicating the situation with a second diffraction grating of period d, located symetrically after the lens.

The propagation of the waves after the lens is obtained directly from eq.3.

$$\Psi(x) = A(x)e^{i\frac{kx^2}{2z}}e^{if(t)} \qquad (4)$$

After traversing a second diffracting grating we obtain:

$$\Psi(x) = A(x)e^{i\frac{kx^2}{2z}}e^{if(t)}e^{\pm i2px\frac{\operatorname{sen}q}{l}} \qquad (5)$$

Two waves as showed in figure 1, $\Sigma_1$ representing a virtual orthoscopic image and $\Sigma_2$ a real pseudoscopic image, which we indicated as a "conjugate" of expression (3) because z is now a negative value.

This is a kind of conjugated wave, giving a pseudoscopic (depth reverted) image of the scene, in three dimensions with continuous parallax. This image was not refered previously on any imaging system and may now be considered as characteristic of diffracting elements. Perhaps, it could be applied in some cases of phase conjugation techniques.

Another application of this kind of images is that of projecting a complete 3D image from a very small optical element, without needing large optics, a result that was not previously achieved by any technique.

A further step is to calculate the position and aberrations of both images This step will be discussed alsewhere, but we show in Figure 2 the corresponding ray path that allows for obtaining a "holoimage" of the object.

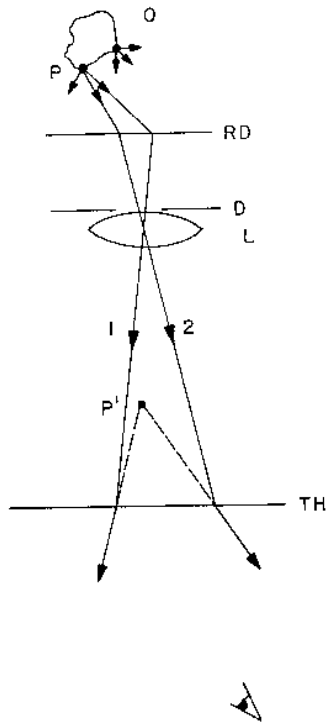

Figure 2

Rays 1 and 2 are of two different wavelengths but, combined, generate the virtual image of the object point P at location P'. This image may be enlarged or reduced by the lens action, because the holographic screen TH that replaced the second diffraction grating allows for that.
This also allows for the enlargement of a common white-light hologram.
According to reference (1) and (2), the illumination of a common off-axis hologram under white light is a situation very similar to the presence of an object, also illuminated under white light, behind a diffraction grating. So that the enlargement and conversion to white light illumination of a common laser hologram are simultaneously obtained by projection on a holographic screen.

Conclusions

We demonstrate that it is possible by using only diffraction processes, to emulate visual results of a holographic image without requiring a registering previous step. This result is very useful for application in visual arts, publicity and for holographic cinema.

Acknowledgments

The author would like to acknowledge finantial help from the Foundation of Assistance to Research of the Sao Paulo State - FAPESP.

References

1) Lunazzi, J.J., Opt Eng. 29, 1 (1990) pp. 9-14

2) Lunazzi, J. J.) Opt.Eng. 29, 1 (1990) pp. 15-18